\documentclass[sn-mathphys,Numbered]{sn-jnl}

\usepackage{graphicx}%
\usepackage{multirow}%
\usepackage{amsmath,amssymb,amsfonts}%
\usepackage{amsthm}%
\usepackage{mathrsfs}%
\usepackage[title]{appendix}%
\usepackage{xcolor}%
\usepackage{textcomp}%
\usepackage{manyfoot}%
\usepackage{booktabs}%
\usepackage{algorithm}%
\usepackage{algorithmicx}%
\usepackage{algpseudocode}%
\usepackage{listings}%
\usepackage{fontenc}%
\usepackage{booktabs}
\newcommand{\tabitem}{~~\llap{\textbullet}~~}

\begin{document}

\title[Article Title]{Enabling Telecommunication Relay Service Research with ACE Omni Platform}


\author*[1]{\fnm{Harrison} \sur{Bourikas}}\email{hbourikas@mitre.org}
\equalcont{These authors contributed equally to this work.}

\author[2]{\fnm{Eric} \sur{Kosinski}}\email{ekosinski@mitre.org}
\equalcont{These authors contributed equally to this work.}

\author*[1]{\fnm{Karina} \sur{Roundtree}}\email{kroundtree@mitre.org}
\equalcont{These authors contributed equally to this work.}

\author[1]{\fnm{Ronna} \sur{ten Brink}}\email{rtenbrink@mitre.org}
\equalcont{These authors contributed equally to this work.}

\author[2]{\fnm{Mike} \sur{Woodman}}\email{mwoodman@mitre.org}
\equalcont{These authors contributed equally to this work.}

\author[3]{\fnm{Vincent} \sur{Ybarra}}\email{vybarra@mitre.org}
\equalcont{These authors contributed equally to this work.}

\affil*[1]{\orgname{The MITRE Corporation}, \orgaddress{\street{202 Burlington Rd}, \city{Bedford}, \postcode{01730}, \state{MA}, \country{USA}}}

\affil[2]{\orgname{The MITRE Corporation}, \orgaddress{\street{1030 Broadway St}, \city{Shrewsbury}, \postcode{07702}, \state{NJ}, \country{USA}}}

\affil[3]{\orgname{The MITRE Corporation}, \orgaddress{\street{1150 Academy Park Loop}, \city{Colorado Springs}, \postcode{80910}, \state{CO}, \country{USA}}}


\abstract{The Telecommunications Relay Service (TRS) industry is comprised of organizations supported and regulated by the Federal Communications Commission that strive to provide deaf, hard of hearing, or DeafBlind individuals with functionally equivalent telecommunication services. Research that assesses current and proposed TRS technologies is used to help close functional equivalence gaps and create recommendations for regulation. TRS researchers spend a considerable amount of time and effort developing experimental environments, which has limited the field’s ability to produce empirical studies. In response to this challenge, the MITRE Corporation has developed a telecommunications research platform called Accessible Communications for Everyone (ACE) Omni. This platform enables researchers to efficiently set up TRS experimental environments, emulate functionality of current TRS technologies, and test new technology solutions. Various design processes, information gathering activities, and the development of personas, research workflows, and functional requirements were leveraged in the design of ACE Omni. A preliminary validation study was conducted via in-lab piloting activities and in vivo to collect real-world TRS user data. Challenges during validation were addressed by making ACE Omni and/or study protocol modifications, and lessons learned are discussed. ACE Omni has the potential to reduce the time and financial costs associated with setting up and running TRS studies, which can promote more TRS research and enable improved service, telecommunication experiences, and outcomes for the community of TRS users.}




\maketitle

\section{Introduction and Motivation}\label{sec1}

The Telecommunications Relay Service (TRS) industry strives to provide individuals who are deaf (individuals who are deaf in the audiological sense and those who identify as culturally deaf) or hard of hearing (DHH) \cite{TRS2024} with functionally equivalent telecommunication services \cite{ITU2018}. The TRS community has debated extensively about what “functional equivalence” means; however, it can be thought of as “the capability to which persons with different range of abilities (in particular, persons with disabilities and persons with specific needs) are able to use a communication service or system with a level of offered functions and convenience-of-use that is similar to those offered to the wider group of users in a population \cite{ITU2018}.” The TRS industry is a multi-billion-dollar industry comprised of several organizations, including service providers and technology developers, that are supported and regulated by government agencies, such as the Federal Communications Commission (FCC).

TRS use has continued to increase and is projected to grow worldwide with an anticipated compound annual growth rate at or above 10\% from 2023 to 2032 \cite{Wadhwani2022}. The growing TRS user base has garnered academic and industry research community interest to evaluate the efficacy and performance of existing and new TRS solutions. However, TRS researchers experience challenges developing experiment environments. Prior to conducting a TRS study, TRS researchers must perform activities including: 

\begin{enumerate}
    \item defining the experimental design;
    \item setting up necessary internet protocols, servers, clients, software, and hardware to connect with and collect data from participants (all of which differ depending on what TRS is being evaluated);
    \item setting up the necessary callers’ physical environments; and
    \item determining logistics and personnel to conduct the study. 
\end{enumerate}

These pre-study activities influence how difficult it will be to conduct the study, aggregate data outputs (e.g., system and participant quantitative and qualitative data), and analyze the data. As experimental designs become more complex, such as involving multiple TRS users, CAs, interpreters, or types of assistive communication technologies, it becomes more difficult, time-consuming, and costly to conduct these studies. Further, many TRS experimental setups have limited reusability from study to study and are rarely shared between TRS researchers. This can arise due to the considerable development, financial investment, and work necessary to recreate unique TRS environments. These challenges reduce the research community’s ability to conduct frequent and timely human-in-the-loop studies.

To address these challenges, the MITRE Corporation\footnote{MITRE, an operator of six federally-funded research and development centers, applies systems thinking across government, industry, and academia to solve whole-of-nation challenges.} (referred to as MITRE), on behalf of the FCC, has developed the open-source Accessible Communications for Everyone (ACE) Omni research platform available on GitHub \cite{Omni}. ACE Omni leverages User Interface (UI) and software design concepts from three TRS applications created for both the research and consumer domains: ACE Quill\footnote{Information about ACE Quill can be found at \cite{Quill}.}, an open-source, simulated Internet Protocol Captioned Telephone Service (IP CTS)/Internet Protocol Relay (IP Relay) platform designed as a research and evaluation tool for captioning and IP Relay systems; ACE Direct\footnote{Information about ACE Direct can be found at \cite{Direct}.}, an omni-channel, universal access, Direct Video Calling call center platform that enables organizations to make/receive video calls to/from anyone and anywhere, and routes those calls to agents fluent in American Sign Language (ASL); and ACE Consumer Access Platform (not published), a Video Relay Services (VRS) platform that lowers barriers-to-entry for parties interested in standing up a VRS system outside of provider offerings. 

ACE Omni enables TRS researchers to replicate the components and functionality of current TRS solutions in a single platform. For each service emulated, TRS researchers can configure, combine, or remove functions or interface elements according to their use cases and can add custom-built functions or elements, allowing them to envision and test new technological solutions. Studies created using ACE Omni can include emulations of multiple types of TRS and multiple participant sessions. ACE Omni can promote more frequent and timely studies that will help ensure TRS users’ needs and challenges are appropriately integrated into the design of existing, updated, or new TRS solutions before they become available to the public, which can bolster usage and improve user satisfaction. All of these benefits will enable improved service and assistive communication technology experiences and outcomes.

\begin{figure}[h]
\centering
\includegraphics[width=\textwidth]{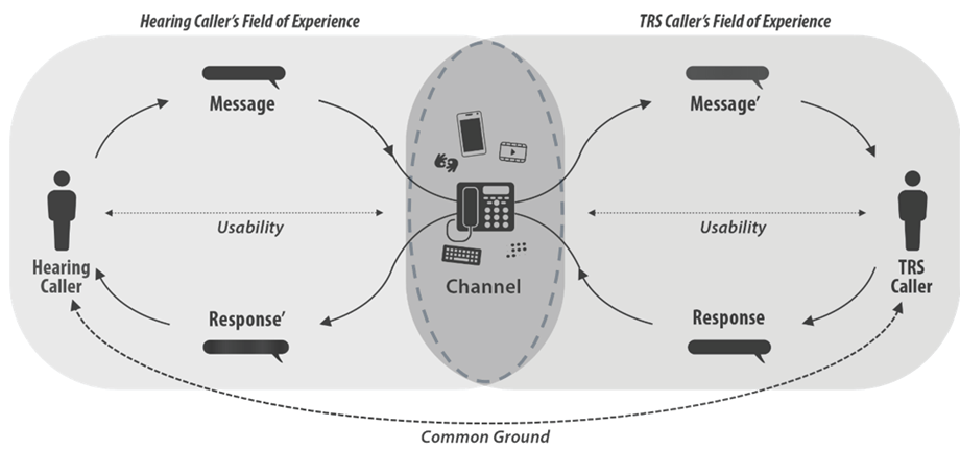}
\caption{The Effective Communication Framework (ECF).}\label{fig1}
\end{figure}

This paper aims to (1) discuss the methods leveraged to design ACE Omni, (2) discuss lessons learned from in-lab piloting activities and during a study evaluation in the field, and (3) provides recommendations to improve ACE Omni for future studies. The Effective Communication Framework (ECF) \cite{Pfaff2025} is leveraged as a means of discussing which elements are influencing the telecommunication conversation and what should be considered when evaluating TRS. Figure~\ref{fig1} shows how the callers (i.e., Hearing and TRS) engage with one another through a Channel that consists of assistive communication technologies, communication assistants (CAs), and/or sign language interpreters for telecommunication. Messages and Responses that are formulated by the Hearing or TRS caller can be manipulated (i.e., Message’ and Response’) if the input communication technology makes the process of expressing a message or response difficult, the message or response changes due to the reliability of the Channel, and the output communication technology makes receiving a message or response difficult. 


\section{Background}\label{sec2}

TRS research has increased within the last decade especially after the introduction of IP CTS and VRS. There are four areas TRS research has primarily focused on: 

\begin{enumerate}
    \item users experience with TRS (Caller element of the ECF);
    \item evaluating and understanding TRS services (Channel element of the ECF), which includes interpreter and CA experiences; 
    \item evaluating existing and novel TRS solutions (various elements of the ECF); and 
    \item designing and evaluating other assistive communication technologies (various elements of the ECF) that can or will interface with TRS. 
\end{enumerate}
 
The methodologies and set ups used in these areas are discussed as a means of understanding what researchers need to perform their studies. 

\subsection{User Experience}\label{sec3}

The primary methods used to collect TRS user experience have been surveys, focus groups, and interviews. For example, a survey was conducted to determine the preference, frequency of use, and accessibility of technologies used by DHH adults in the United States \cite{Maiorana-Bases2014}. The study found that smartphones and computers were more frequently used compared to other telecommunication devices and that the majority of people used technology at home to check email, text message, surf the Internet, attend videoconferences, and write documents. Other participants shared their VRS experiences via focus groups \cite{Brunson2010}. The VRS users experienced misunderstandings due to the interpreter’s inability to understand finger spelled words; waiting several minutes to connect with interpreters, likely due to the limited number of interpreters and high demand for their services; and difficulties correcting interpreters when they signed something incorrectly. These participants employed strategies when challenges arose, such as terminating the call and trying again later. 

Other challenges have occurred when DHH participants, who primarily use ASL to communicate, participate in meetings remotely with hearing individuals. For example, one group used an audio bridge to connect meeting attendees \cite{Vogler2013}. The DHH participants were unable to see one another and relied on their VRS interpreters to convey information. This set up caused high levels of frustration and ultimately reduced the DHH attendees’ participation in meetings or they stopped attending the meetings altogether. While some of these challenges have been addressed with the advent of new technology, there are still challenges DHH persons experience. For example, a study conducted semi-structured interviews, via Zoom, to determine what challenges DHH participants had with authentication in customer service calls \cite{Andrew2023}. Thematic analysis unveiled negative consequences when a third party was required to authenticate the user. Those included: the burden of finding a third party, compromised user privacy and security due to needing to share personal information with others, and in some cases customer service representatives refused to work with a third-party mediator.  

The Australian Communication Exchange Ltd evaluated, via surveys and focus groups, the CTS Australian trial’s impact on health-related quality of life and wellbeing for those who have a hearing deficiency \cite{Connelly2011}. In general, access to a captioned telephone reduced users’ feeling upset or embarrassed by their hearing problem when using a phone and getting frustrated while communicating on the phone. Positive effects were observed for those that had the opportunity to use a captioned telephone in their workplace, including improved work satisfaction; communication with colleagues, coworkers, and customers; and range of tasks they could perform at work. The two sources of dissatisfaction were associated with the limited hours of captioning service operation and the technology setup. 

More formalized interview protocols, involving various tasks, knowledge elicitation, and table-top exercises, have been used to understand IP CTS users’ cognitive demands and abilities to effectively communicate \cite{Urqueta2023}. The strategies these IP CTS users leveraged were adapting their communication behaviors (e.g., asking clarifying questions or asking the other participant to be patient while captions arrived); selecting technology that best suited specific situations (e.g., using particular providers and hardware for different types of calls); and disclosing their hearing loss, which caused user discomfort and worry of being misperceived as less capable. This evaluation and the other user experience research described emphasize the importance of understanding what elements negatively impact TRS user experience to improve current and future assistive communication technologies and services.

\subsection{Service Evaluations}\label{sec4}

Surveys, conducted through SurveyMonkey and semi-structured interviews, were used to understand American \cite{Bower2015} and Canadian interpreter \cite{Chang2022} experiences working in VRS environments. Common interpreter stressors found across these studies were managing emotional calls, especially when callers were angry; work culture expectations (i.e., answering as many calls as possible, short times between calls, and balancing providing interpreting services with other job responsibilities); and interpreting calls with limited contextual information. Constantly thinking about mechanisms to manage interactions between callers and establishing caller attendee roles also caused interpreter stress. These strategies ranged from behaving as a conduit to a co-provider/co-creator. Two studies assessed interpreter interactions: one used authentic calls by Swedish’s VRS (Bildtelefoni.net) \cite{Warnicke2012} and the other simulated non-emergency police VRS calls through SignVideo’s Scottish branch \cite{Skinner2023}. It was observed that interpreters hold the power to sanction caller turn taking through visible (e.g., placing their hand in a location, nodding their head, positioning their body in different orientations), audible (e.g., rendering an utterance), and rendition (e.g., modify what was expressed to buy time) strategies \cite{Warnicke2012}. The necessity to manage interactions also depends on the context of the call. Interpreters who assumed a broader role (i.e., not just a translator) in non-emergency police VRS calls resulted in officers having a better understanding of VRS, the purpose of the call, and the roles each attendee had \cite{Skinner2023}. 

Various studies have recommended ideas for reducing interpreter stress, such as reducing call volume, increasing break time, providing more support opportunities, improving policies, and practicing breathing strategies. One study evaluated how interpreters’ participation in an adapted Demand Control-Schema case conferencing group helped reduce stress and attrition in VRS \cite{Wilbert2021}. The focus groups and interviews uncovered that participants found the conferencing group to be effective in reducing stress because they were able to gain tools to manage calls and consider prospectives from other community members. Support strategies, such as teaming practices, have been introduced at VRS call centers to aid interpreters in need \cite{Rainey2013}. Interpreters often needed help when they couldn’t understand the callers, there were technological difficulties, or the call topic was complex (e.g., legal, 911, job interviews). 

The emphasis of working conditions in VRS environments has been intensified as more interpreters transition from on-site interpreting environments to remote at-home interpreting. Three studies, using questionnaires, surveys, focus groups, and interviews, examined the experiences of interpreters transitioning from on-site to remote interpreting during and post-Coronavirus disease. While some positive consequences arose with the transition to remote interpreting, such as support with ways to deliver interpreting services, new software platforms, more time between jobs, and ease with filling schedules \cite{Roman2023}, negative consequences also occurred. Physical and mental health concerns emerged and interpreters needed to acquire new skills (e.g., coping with the multimodal nature of online interpreting) and manage their workload \cite{DeMeulder2024}. To mitigate these challenges and improve perceptions, it was suggested that sufficient managerial support, training, materials, resources, and experience remotely interpreting before committing to a full-time job are needed \cite{Roman2022}.

\subsection{Existing and Novel TRS Solutions}\label{sec5}

Authentic VRS recordings were used to investigate how callers used Sweden’s VRS text-function and how it influenced their interactions \cite{Warnicke2021}. Conversation Analysis was conducted to analyze the used VRS interactional resources, including: signed and/or spoken language, written text, gaze, facial expressions, body movement, and gestures. The added text-function in VRS proved to be a valuable as it helped to:

\begin{enumerate}
    \item conduct repairs in the conversation,
    \item prevent possible misunderstandings, 
    \item facilitate the progression of the interactions, and 
    \item overcome language differences. 
\end{enumerate}

Another study conducted interviews to understand officer, VRS provider, and DHH caller perceptions using a non-emergency VRS system \cite{Skinner2021}. In general, participants felt that interpreters needed to assume a broader role to ensure successful caller interactions due to the lack of knowledge or preparedness from the officers. DHH callers lacked trust and confidence that the interpreters and police would be able to assist them and viewed the VRS platforms as inadequate and cumbersome to use because of required apps or plug-ins. These studies emphasize the importance of evaluating how ECF elements interact with one another and influence users’ experiences.  

A videoconferencing setup using FuzeMeeting to accommodate DHH meeting attendees, who were originally using an audio bridge in their meetings, was developed as an effort to improve users’ experiences with ECF elements \cite{Vogler2013}. The system’s requirements, identified by the DHH users, included a minimum video frame rate of 20-30 frames per second to support sufficient video quality for six hours of continuous meetings, no additional hardware or software to function, and was portable and easy to set up. This solution is similar to videoconferencing platforms where almost every meeting attendee has their own respective monitor and webcam so that all participants can see remote and local attendees. The setup changes resulted in a 50\% higher DHH member attendance. Another study also leveraged user feedback to design a high-fidelity VRS mobile interface \cite{Kushalnagar2016}. An initial survey was used to capture participants’ feedback about preliminary prototypes and yielded suggestions such as adding call time, multiple views, separation of call history, and being able to choose functions from call settings. The feedback was then implemented into secondary prototypes, followed by a final survey. The final feedback was relatively positive (i.e., the prototypes were clear to understand); however, individual differences were mentioned, such as disagreement with the color schema and contrast.

Other studies have developed video and caption telecommunication solutions, but did not incorporate user evaluations of the systems. For example, one study added the GStreamer2 Multimedia Framework to a Linphone (i.e., a Session Initiate Protocol-based Voice Over Internet Protocol Phone) desktop platform’s Media Streamer 2 Library \cite{Ongtang2022}. The client-side implementation enabled multi-party flexible audio, video, and text communication with recording capability, all of which are critical VRS functions, and did not require the client to install additional software on their machine, helping reduce server resource consumption. Another study developed a VRS prototype using the open-source WebRTC platform with JavaScript so that the platform could run on Android-operating mobile devices by accessing a WebRTC enabled web-browser (e.g., Firefox or Chrome) \cite{Henney2019}. Three main components of the system included: the MediaStream Application Programming Interface (API), which allowed the browser access to local input and output devices; the RTCPeerConnection API, which set up a secure peer-to-peer connection; and the RTCDataChannel API, which allowed for secure bi-directional data exchange between peers. Another study also focused on mobile devices by developing an all-in-one solution of VRS, third party video calls, voice to text caption, and NextGen 911 \cite{Behm2021}. Their prototype used WebRTC technology that first checked operating system settings to see if accessibility settings were enabled and then handled VRS call flow in coordination with an enhanced over the top video service core. The prototype modified the native mobile dialer app to send audio stream from the call to the chosen IP CTS provider and then the captioned text was displayed on the mobile screen. 






\subsection{Assistive Communication Technology Research Applicable to Current and Future TRS}\label{sec6}

Other research has focused on assistive communication technologies that may be applicable to future TRS. Some studies have explored Video Relay Interpreting (VRI) challenges \cite{Mauldin2022}, investigating user satisfaction with VRI services in health care settings \cite{Kushalnagar2019}, and evaluating VRI implementation in hospitals \cite{Marshall2019}. VRI can be similar to VRS, therefore these interpreter and user experiences may relate to experiences with VRS. Other studies focused on signing solutions that incorporate avatars. For example, exploring the cultural acceptance of signing avatars instead of human interpreters \cite{Othman2024} and rendering an avatar signing on augmented reality glasses \cite{Guo2023}. Both solutions could influence future VRS, transitioning away from human interpreters. 

Some studies were interested understanding requirements needed to create inclusive videoconferencing platforms when participating via signed languages \cite{Ang2022} and to provide improved accessibility on teleconferencing platforms \cite{Beldon}. Many captioning solutions have focused on Automatic Speech Recognition (ASR) \cite{Chotimongkol2024}, which is leveraged in IP CTS. For example, a study investigated the benefits ASR offered when captions provided to the caller were manipulated as well as the caption word accuracy levels \cite{Zekveld2009}. Others explored improving ASR using supervised natural language processing models to learn the importance of some words versus others \cite{AlAmin2023}. Providing real-time vibration feedback on participants’ wrists has also been proposed to help participants perceive the number of speakers and speakers’ ages and genders \cite{Wang2023}. One study even examined how providing captions to hearing partners of what they said influenced conversational accuracy and dynamics \cite{Kang2024}. These investigations may influence future TRS functionality and how other platforms engage/incorporate TRS. 






\section{Methods}\label{sec7}

The conceptualization and development of ACE Omni leveraged methodologies, such as Design Thinking \cite{DesignThinking2016}, Agile Methods \cite{Silva2011}, and discussions with Subject Matter Experts (SMEs) in the telecommunication services domain. Design Thinking was used to understand the ECF’s TRS Caller’s Field of Experience and Channel, which included the needs, barriers, challenges, goals, and experiences of TRS users and researchers. TRS user and researcher personas, detailing a generalizable TRS research workflow, were created to guide the development of functional requirements and design of low-fidelity mockups. These artifacts were substantiated by a TRS literature review, an assessment of TRS providers and their technologies, a market survey on sustainable practices of existing research platforms used for human-subjects research, and interviews with TRS research SMEs. Once the problem space was conceptualized and findings synthesized, initial design concepts were developed and iterated on. MITRE software developers used Agile Methods to implement a minimum viable product and prioritized requirements that accommodated MITRE researchers who were going to conduct TRS research at an upcoming conference (discussed more in Section \ref{sec14}). 

\subsection{Information Gathering}\label{sec8}

Various methods were leveraged to gather information relative to ACE Omni development including conducting a TRS literature review (discussed in Section \ref{sec2}), a market survey, discussions with SMEs and TRS providers, as well as attending conferences. 

A market survey was conducted on various research platforms to inform how to develop a sustainable ACE Omni. Sustainability is valuable because it allows researchers to improve the “reproducibility and reusability of research” \cite{Manifesto}. Ten platforms used for human-subjects research were evaluated using the Software Sustainability Institute’s 18-criterion assessment for sustainability \cite{Jackson2011}. Results of the market survey indicated that the research platforms have sustainable and usable tools for survey and cognitive psychology data collection, but do not include functions frequently used by DHH technology researchers (e.g., captioned calls). Additionally, no platform satisfied all 18 sustainability measures; however, some platforms were largely sustainable and served as a design model for ACE Omni (e.g., PsychoPy \cite{Peirce2019}).

TRS user experiences were collected in collaboration with SMEs at two academic institutions: Gallaudet University (GU) and Rochester Institute of Technology National Technical Institute for the Deaf (RIT/NTID). MITRE designers held conversations about TRS persona development with RIT/NTID, incorporating suggestions related to context and persona use cases (e.g. users experience technological issues or updates that interrupt TRS usage). MITRE designers also leveraged information from a GU presentation on VRS, IP CTS, and CTS operations user experiences.

MITRE requested usage information from current TRS Providers to understand TRS usage trends (e.g. usage minutes per month per service type), and garner insight into who uses TRS and why (i.e., demographics and motivations). MITRE synthesized TRS Provider responses, gaining a holistic view of TRS usage and lending context and statistics-based details to the TRS user persona set.

TRS user experiences, common barriers, and challenges with current technology solutions were also identified while attending the Axe-con 2023 conference. The conference’s focus was on companies leading large-scale accessibility and best practices across design, development, management, testing, and legal efforts. “The User Research and Personas for BARD Mobile” \cite{Deque2023} presentation by the National Library Service for the Blind and Print Disabled, a service of the Library of Congress, highlighted considerations for individuals who are partially sighted or blind.

\subsection{Personas and Workflows}\label{sec9}

The information collected in Section \ref{sec7} was used to create TRS user and researcher personas, as well as inform functional requirements. MITRE designers developed 29 personas characterizing a spectrum of TRS users, how they interact with TRS, and the barriers and challenges they face when using TRS. This enabled MITRE designers to understand who comprises the TRS user community, how their differences lead to technology and service needs, and how to qualitatively evaluate how TRS serve users in different scenarios. The MITRE designers used the persona set to characterize who TRS researchers might recruit as study participants and detail the many complex scenarios and technology ecosystems TRS researchers might try to replicate.

\begin{figure}[h]
\centering
\includegraphics[width=\textwidth]{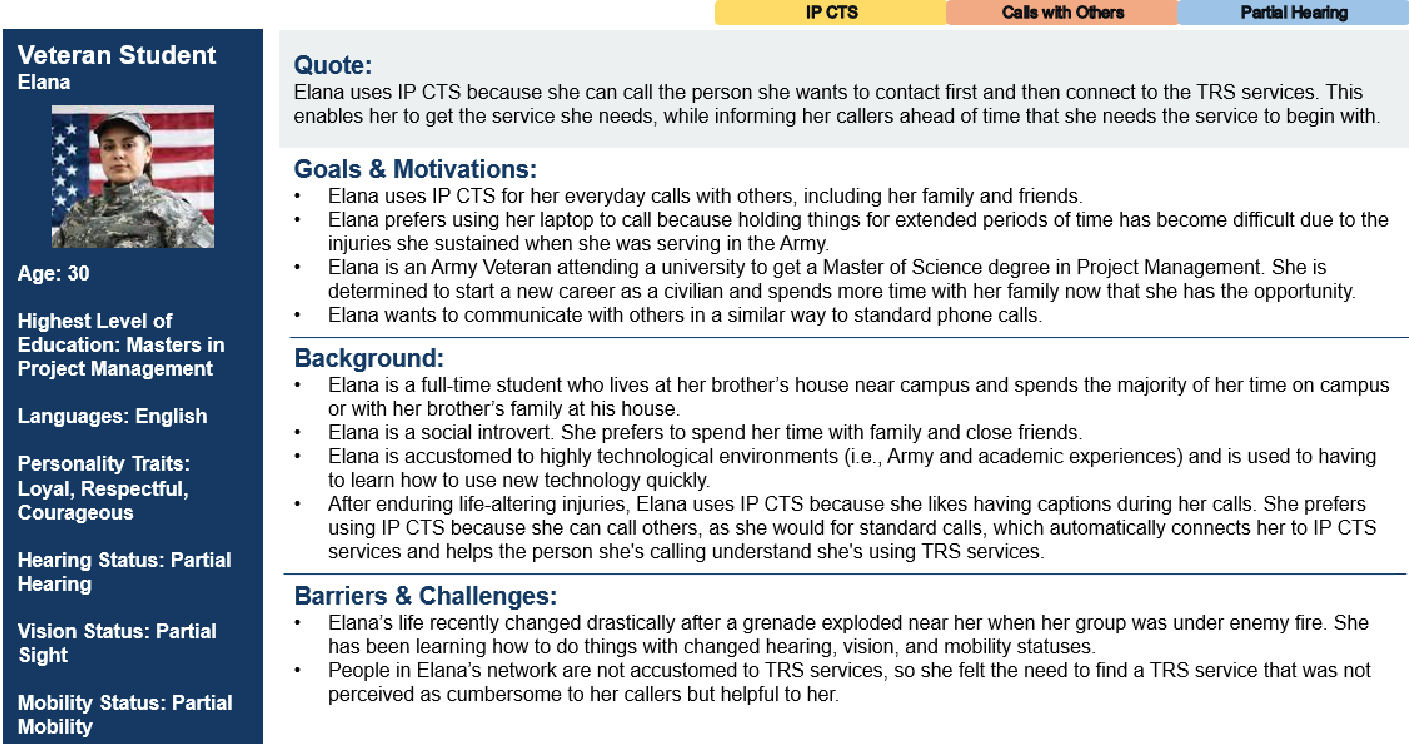}
\caption{Example of TRS user persona.}\label{fig2}
\end{figure}

Figure~\ref{fig2}, for example, is TRS user persona Elana and includes information about her profession; age; education level; languages she uses; personality traits; and her hearing, vision, and mobility status information (Elana has partial hearing, sight, and mobility). In the upper-right quadrant, a classification scheme is included to describe the TRS used (e.g., IP CTS), a typical use case consistent with the persona’s background (e.g., calls with others), and their hearing status. The user’s goals and motivations (e.g., Elana wants to communicate with others in a similar way to standard phone calls), their background (e.g., Elana is a full-time student), as well as barriers and challenges they face (e.g., People in Elana’s network are unaccustomed to TRS so she uses a TRS that is perceived as less cumbersome to her callers) are detailed. 

A generalized TRS Research Workflow, in Table \ref{tab1}, was created to understand the pre-, post- and experiment TRS research phases and to inform TRS researcher personas. The phases included: designing the experiment (Phase 1), developing the technical setup (Phase 2), executing the experiment and collecting data (Phase 3), and reporting findings (Phase 4). The processes that could be supported by a TRS-specific research platform were identified and capabilities to support those processes were defined and translated into functional requirements (detailed in Section \ref{sec10}).

\begin{table}[h]
\caption{Generalized TRS research workflow.}\label{tab1}
\begin{tabular*}{\textwidth}{@{\extracolsep\fill}cccc}
\toprule%
\multicolumn{2}{@{}c@{}}{Pre-Experiment} & \multicolumn{1}{@{}c@{}}{Experiment} & \multicolumn{1}{@{}c@{}}{Post-Experiment} \\\cmidrule{1-2}\cmidrule{3-3}\cmidrule{4-4}
Phase 1: Design & Phase 2: Develop & Phase 3: Execute & Phase 4: Report \\
Experiment & Technical Setup & Experiment and & Findings \\
 &  & Collect Data & \\
\midrule
\multicolumn{1}{@{}l@{}}{\tabitem Consider technology}  & \multicolumn{1}{@{}l@{}}{\tabitem Define experiment} & \multicolumn{1}{@{}l@{}}{\tabitem Experiment} & \multicolumn{1}{@{}l@{}}{\tabitem Clean data} \\
\multicolumn{1}{@{}l@{}}{and logistics for} & \multicolumn{1}{@{}l@{}}{design requirements} & \multicolumn{1}{@{}l@{}}{environment is} & \multicolumn{1}{@{}l@{}}{\tabitem Analyze data} \\
\multicolumn{1}{@{}l@{}}{experiment/research} & \multicolumn{1}{@{}l@{}}{\tabitem Consult technical} & \multicolumn{1}{@{}l@{}}{operational} & \multicolumn{1}{@{}l@{}}{\tabitem Write report} \\
\multicolumn{1}{@{}l@{}}{\tabitem Investigate previous} & \multicolumn{1}{@{}l@{}}{experts} & \multicolumn{1}{@{}l@{}}{\tabitem Follow experiment} & \multicolumn{1}{@{}l@{}}{and publish findings} \\
\multicolumn{1}{@{}l@{}}{research} & \multicolumn{1}{@{}l@{}}{\tabitem Create experiment} & \multicolumn{1}{@{}l@{}}{protocol (designed in} & \multicolumn{1}{@{}l@{}}{\tabitem Defend findings} \\
\multicolumn{1}{@{}l@{}}{\tabitem Establish research} & \multicolumn{1}{@{}l@{}}{technical requirements} & \multicolumn{1}{@{}l@{}}{Phase 1 and enabled} & \multicolumn{1}{@{}l@{}}{\tabitem Use findings to} \\
\multicolumn{1}{@{}l@{}}{questions and} & \multicolumn{1}{@{}l@{}}{\tabitem Install, configure,} & \multicolumn{1}{@{}l@{}}{in Phase 2) with} & \multicolumn{1}{@{}l@{}}{jumpstart new} \\
\multicolumn{1}{@{}l@{}}{independent/dependent} & \multicolumn{1}{@{}l@{}}{and test experiment} & \multicolumn{1}{@{}l@{}}{participants} & \multicolumn{1}{@{}l@{}}{research experiments} \\
\multicolumn{1}{@{}l@{}}{variables} & \multicolumn{1}{@{}l@{}}{setup} & \multicolumn{1}{@{}l@{}}{\tabitem Monitor incoming} &  \\
\multicolumn{1}{@{}l@{}}{\tabitem Get study approval} & \multicolumn{1}{@{}l@{}}{\tabitem Train research team} & \multicolumn{1}{@{}l@{}}{data} &  \\
\multicolumn{1}{@{}l@{}}{\tabitem Pilot experiment} & \multicolumn{1}{@{}l@{}}{on the experimental} & \multicolumn{1}{@{}l@{}}{\tabitem Ensure data is} &  \\
\multicolumn{1}{@{}l@{}}{\tabitem Recruit participants} & \multicolumn{1}{@{}l@{}}{technical setup} & \multicolumn{1}{@{}l@{}}{deidentified} &  \\
\botrule
\end{tabular*}
\end{table}

Three personas were developed that characterized a TRS research team: a lead researcher; a supporting researcher; and two technical experts with software engineering backgrounds and knowledge in network, server, database, and hardware integration. It was assumed that each persona was an expert in their field with no overlap between roles and responsibilities, which enabled the creation of distinct barriers, challenges, needs, and goals. The research workflow details where and when the three personas contribute their expertise, providing insight into their roles, responsibilities, actions, and collaboration opportunities. For example, a lead researcher leads the design of experiment efforts (Phase 1), which includes defining research questions, conditions, stimuli, and drafting the study protocol. That lead researcher can coordinate with technical experts to create the experiment environment used during research sessions (Phase 2). Supporting researcher(s) can help the lead researcher execute the study, collect data (Phase 3), analyze the data, and report findings (Phase 4).

\subsection{Functional Requirements}\label{sec10}

The findings from Sections \ref{sec8} and \ref{sec9} were used to develop ACE Omni functional requirements. The requirements were developed with respect to four standard system functions and core capabilities groups. Installation refers to the ability to install ACE Omni on most server types, whether it's a local server like a laptop, cellphone, private server, or a cloud-based server such as Amazon Web Services (AWS) EC-2 instance. This hosting flexibility allows TRS researchers to collaborate with participants across multiple locations. The TRS interface setup should enable TRS researchers to create, configure, and save TRS experiences for use in human-subjects experiments. TRS researchers should be able to select data to collect and view collected data. Participant experience enables participants to interact within the parameters of an experiment.

\section{Design}\label{sec11}

ACE Omni (an ECF channel technology) is an open-source, browser-based platform that enables TRS research. ACE Omni can be installed on any server, allowing for flexibility in where TRS researchers host the system and enabling TRS researchers to work with participants in multiple locations. ACE Omni includes VRS and IP CTS human-subjects research tools and enables TRS researchers to augment the existing code to configure ACE Omni beyond current platform capabilities. The platform implements a modular architecture where TRS researchers can configure properties (e.g., caption delay) of independent modules that emulate capabilities of existing TRS to test study conditions per their experiment design. The platform enables TRS researchers to:

\begin{enumerate}
    \item create and configure a study with one or more modules;
    \item pilot, edit, and finalize the study; 
    \item deploy the study to collect data; and 
    \item export the data in a usable format (i.e., comma-separated values (CSV)).
\end{enumerate}

\begin{figure}[h]
\centering
\includegraphics[width=\textwidth]{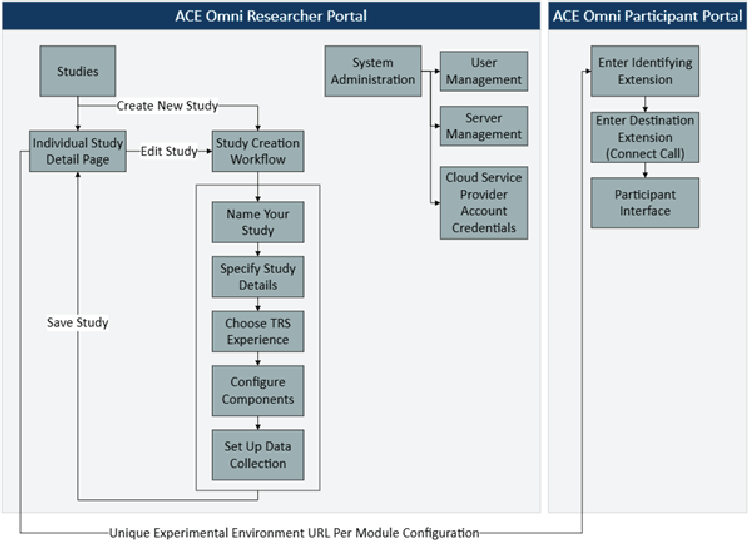}
\caption{ACE Omni sitemap.}\label{fig3}
\end{figure}

ACE Omni has two portals: the Researcher Portal enables TRS researchers to produce TRS experiment environments, administer the system, and view and export collected data; and the Participant Portal enables participants to access experiment environments. Figure~\ref{fig3} illustrates the relationship of pages (gray boxes) contained in the Researcher Portal and Participant Portal. The Researcher Portal includes pages to create new studies or access existing studies. The study creation process chronologically guides (follow the arrows) the TRS researcher through steps including the name of the study, specifying study details, choosing the TRS experience, configuring components, and setting up the data collection. Information of studies can be edited via the study detail page, while platform users and servers can be managed and cloud service provider (CSP) credentials entered via the system administration page. The Participant Portal includes pages for participants to enter identifying extensions that map to experiment environments for individual participants, enter destination extensions to connect calls between participants, and access experiment interfaces.

ACE Omni emulates TRS, which eliminates the need for TRS researchers to use existing TRS solutions that they may not be able to access, be permitted to use, or require stringing together systems with complex interoperability requirements. ACE Omni was developed with a ReactJS \cite{React} frontend, because of the framework’s ability to create multiple small components, which allows for quick implementation and modification. The backend implements four components to enable browser-based emulation, including an Express.js \cite{Express} backend, MongoDB server \cite{MongoDB}, a NodeJS server \cite{NodeJS}, a SocketJS connection, and a DevExtreme library \cite{DevExtreme}; commonly referred to as a MERN Stack \cite{MERN}. The MongoDB server tracks user profiles for registration and identification, and it manages attributes of individual studies. The Express.js acts as the model of the framework and handles communications between the frontend and several of the main components, including selected UI components implemented from the DevExtreme library to support system functions and core capabilities. The SocketJS connection is built into the NodeJS server and is used to establish communication ‘rooms’, which enables TRS researchers to verify each participant has successfully entered their identifying extension before connecting to a call. A WebRTC library \cite{WebRTC} establishes peer-to-peer connections and handles backend call logic.

\subsection{Researcher Portal}\label{sec12}

The Researcher Portal was designed to support actions and components typical of TRS research, based on findings and insights from Section \ref{sec9}. ACE Omni supports the four research phases in Table \ref{tab1} via its Studies page, where TRS researchers create a study to correspond with a research protocol. TRS researchers who have defined their research protocol can translate those details and specifications into the ACE Omni platform using the Create New Study process.

The Researcher Portal offers TRS researchers the ability to manage platform users and servers, and to enter CSP credentials. TRS researchers can create new users and delete existing users in ACE Omni, enabling multiple contributors to access the platform and edit created studies via separate accounts. Upon installation, one default user account is created, which cannot be deleted. TRS researchers can enter and modify CSP account credentials, including keys, logins, passwords, tokens, and certificates, which are used in some of the platform’s TRS emulation modules. Two speech-to-text engines from two CSPs are currently supported: Google \cite{Google} and IBM Watson \cite{Watson}.

\begin{figure}[h]
\centering
\includegraphics[width=\textwidth]{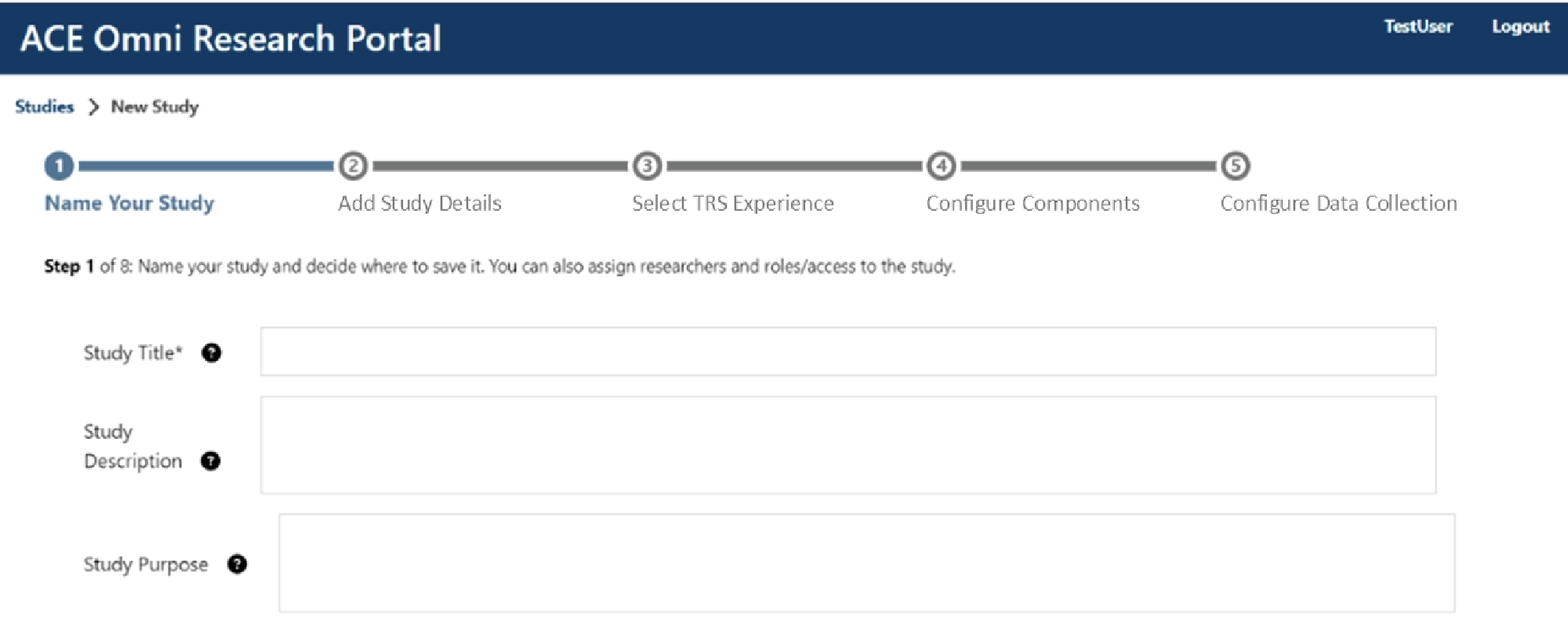}
\caption{Create New Study.}\label{fig4}
\end{figure}

ACE Omni enables TRS researchers to create studies that contain configurable modules that emulate existing TRS, functions to manage study details, and options for collecting and exporting data. When TRS researchers create a study, they are directed to a New Study page (see Fig.~\ref{fig4}) that includes the steps in Table \ref{tab2}. Steps can be edited during or after initial set up, which provides flexibility to TRS researchers create, configure, pilot, and finalize their study before deployment. The New Study page includes a step counter at the top of the page, a content area where TRS researchers enter information, and a row that enables TRS researchers to proceed/go back to the next/previous step, save the study, or cancel to exit the process.

\begin{table}[h]
\caption{Steps of the New Study page.}\label{tab2}
\begin{tabular*}{\textwidth}{@{\extracolsep\fill}ll}
\toprule%
\textbf{Step Number and Name} & \textbf{Enables TRS Researchers To:} \\
\midrule
\#1: Name your study & Record the study’s title, description, and purpose. It captures\\
 & high-level details about a study, including how the study is\\
 & designed and what the TRS researchers are investigating. The \\
 & Study Title, in combination with other metadata, is used to \\
 & generate unique Participant Portal links corresponding to \\
 & experiment configurations.\\
\#2: Add study details & Document the study’s independent and dependent variables,\\
 & research question(s), population size and description, as well as\\
 & planned start and end dates. This step is optional; however, it’s\\
 & intended to capture experiment design information.\\
\#3: Select TRS experience & Select and add one or more default modules (i.e., templates) to\\
 & their study.\\
\#4: Configure components & Manipulate modules by configuring their properties using the\\
 & Configurations panel and corresponding property options.\\
\#5: Configure data collection & Select data elements to collect for each module they add to a\\
 & study. Data collection options are unique to each module, but\\
 & generally include video recordings, audio recordings, screen\\
 & recordings, and captioning data, where applicable.\\
\botrule
\end{tabular*}
\end{table}

ACE Omni includes modules emulating VRS and an ASR-based version of IP CTS. Users can develop and add unique modules via supplemental code, leveraging ACE Omni’s open-source practices. ACE Omni implements default modules that TRS researchers select and customize within a study. A module is a self-contained and configurable experiment environment of a TRS-emulated communication experience. It includes communication affordances for all conversation participants (e.g., DHH or hearing participant) and can emulate an existing or custom TRS experience. TRS researchers can create multiple configurations for one or more modules within a study.

Modules are composed of one or more UIs containing communication affordances for conversation participants. For example, the VRS module includes three UIs presented to the DHH participant, to the CA, and to the hearing participant. UIs consist of one or more properties, which are items that TRS researchers can configure individually to align with sets of conditions. Properties can be a defined value (independent variable), be measurable (dependent variable), or be included/excluded (where permitted) from a UI. For example, the IP CTS module includes properties for caption presentation features (e.g., show/hide speaker labels, left/right/center justification), outgoing audio stream distortion (e.g., type of background noise, duration of simulated packet drops), outgoing audio stream filters (e.g., high-pass versus low-pass filters), and other interface characteristics. UIs are automatically assigned extension numbers as TRS researchers add new module configurations to their study. Participants access UIs corresponding to a particular condition using the appropriate extension. 

\begin{figure}[h]
\centering
\includegraphics[width=\textwidth]{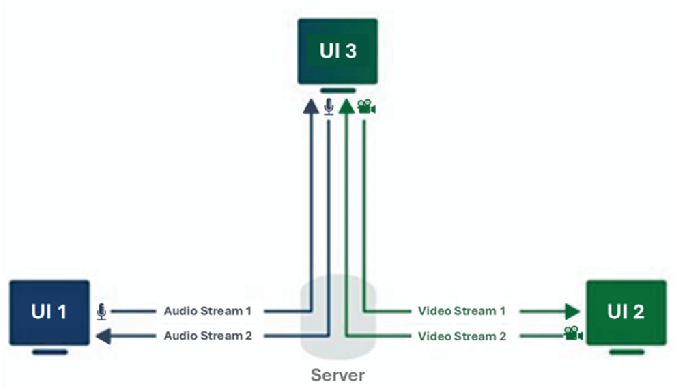}
\caption{VRS module technical diagram.}\label{fig5}
\end{figure}

A technical VRS module diagram illustrated in Fig.~\ref{fig5} demonstrates the operationalization of the ECF. UI 1 is intended to be used by hearing participants, UI 2 is intended to be used by DHH participants, and UI 3 is intended to be used by CAs. Audio Stream 1 data flows from UI 1 to UI 3, where the CA translates spoken words from the hearing participant into sign language. Video Stream 1 data flows from UI 3 to UI 2, where the DHH participant views video of the CA signing. Video Stream 2 data flows from UI 2 to UI 3, where the CA translates sign language from the DHH participant into spoken words. Audio Stream 2 data flows from UI 3 to UI 1, where the hearing participant hears the CA speaking. The three UIs of the VRS module have video and audio properties that TRS researchers configure in Step 4.

\begin{figure}[h]
\centering
\includegraphics[width=\textwidth]{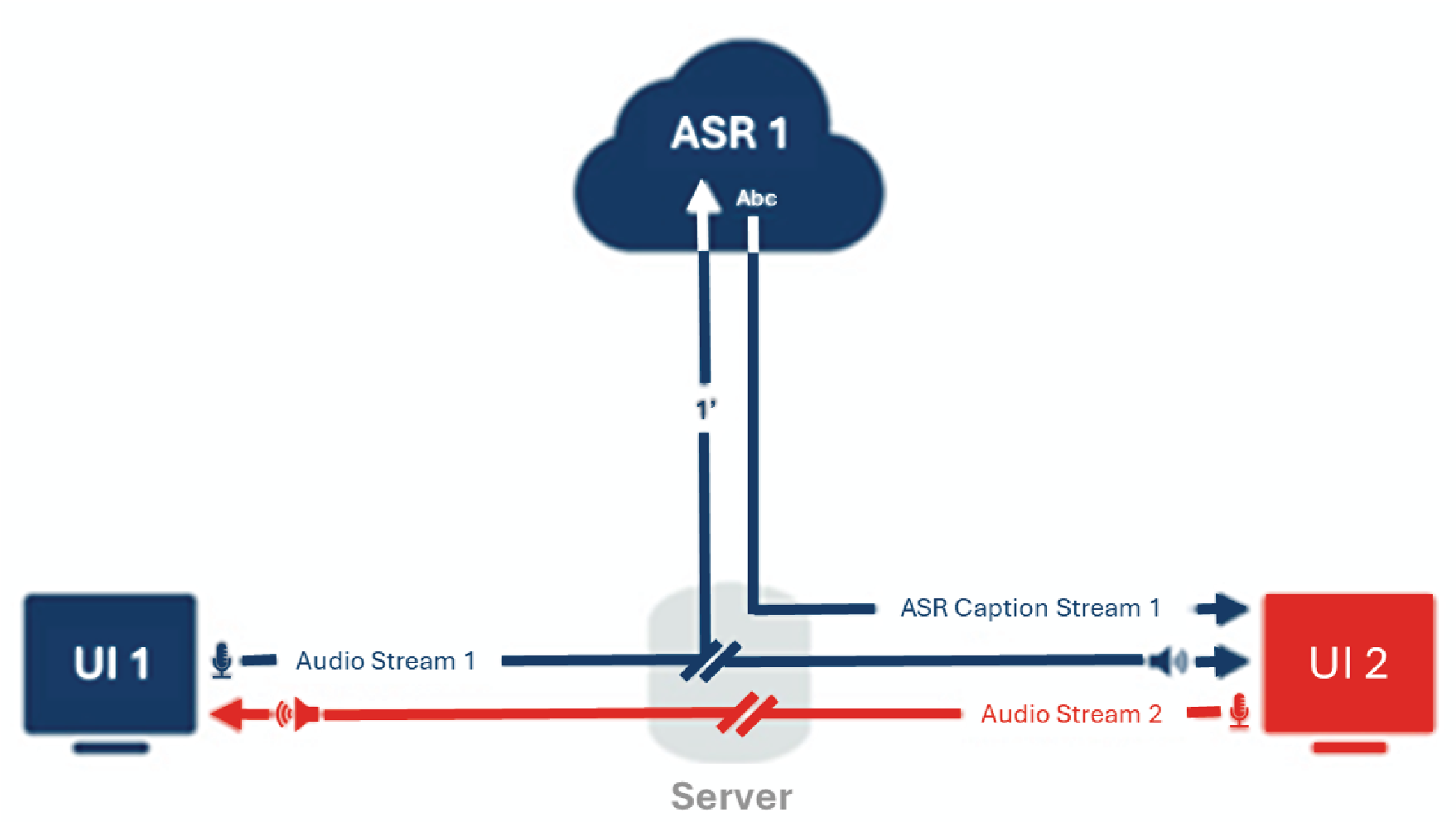}
\caption{ASR-based IP CTS module.}\label{fig6}
\end{figure}

The IP CTS module includes two UIs (see Fig.~\ref{fig6}): UI 1 is intended to be used by hearing participants and UI 2 is intended to be used by DHH participants. Audio Stream 1 is produced by the hearing participant and is received by the DHH participant. TRS researchers can manipulate Audio Stream 1 properties at the microphone, which affects the captions produced by ASR 1 and the audio that the DHH participant receives, or they can manipulate properties at the speaker, which only affects the audio that the DHH participant receives. When Audio Stream 1 passes through the Server, it is divided into Audio Stream 1 and 1’, which enables the module to handle audio to ASR 1 and to UI 2 separately. ASR 1 converts Audio Stream 1’ into ASR Caption Stream 1, which the DHH participant views using UI 2.

\begin{figure}[h]
\centering
\includegraphics[width=\textwidth]{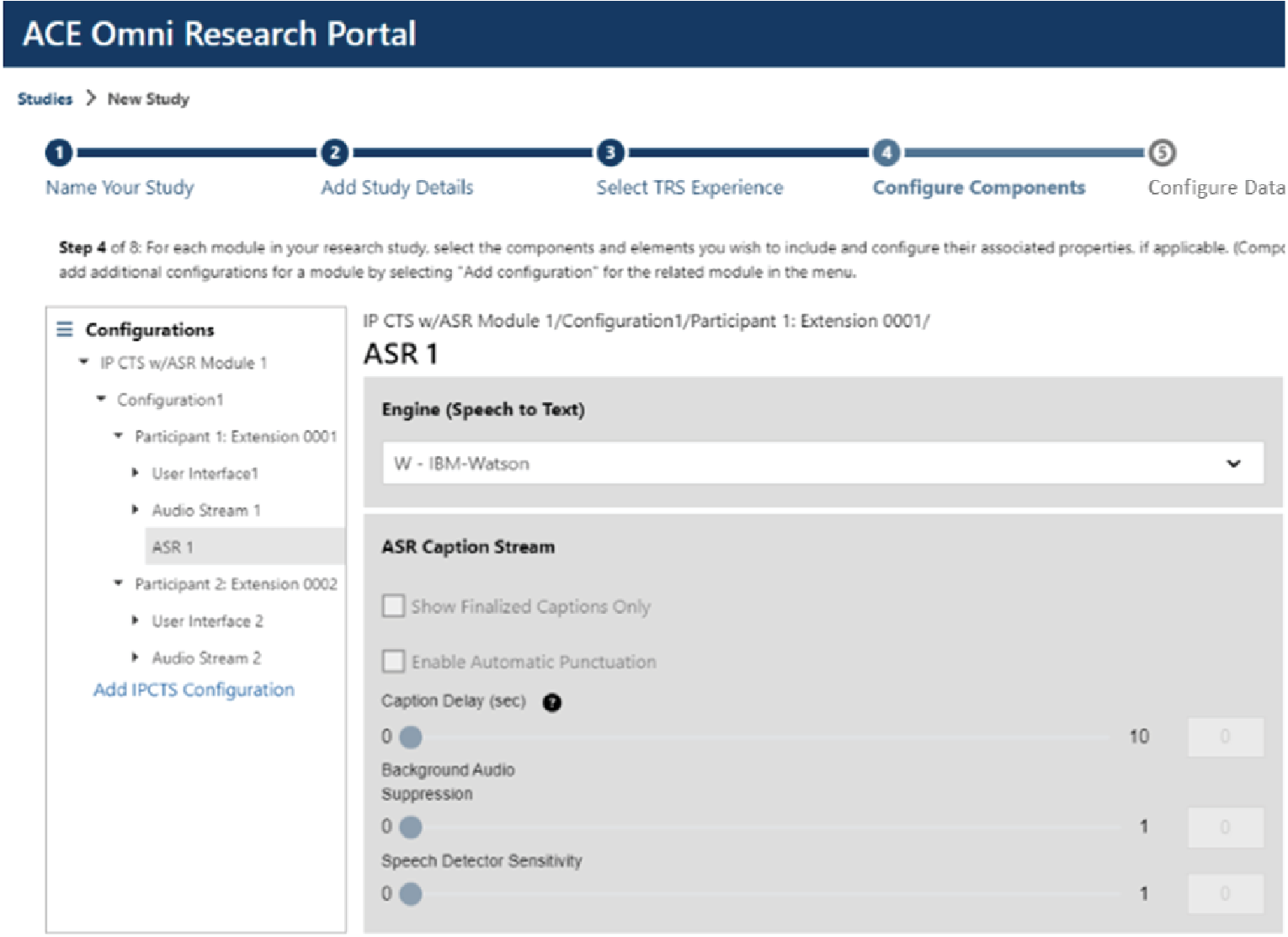}
\caption{Configure Components.}\label{fig7}
\end{figure}

Each module is accompanied by their corresponding technical diagram that details module components and how data moves between UIs. A configuration is automatically added by the system in the Configurations panel (see Fig.~\ref{fig7}) per module, matching selections made in Step 3. TRS researchers can add any number of configurations to correspond with conditions of their experiment design. 

\subsection{Experimental Environments and the Participant Portal}\label{sec13}

Once TRS researchers create and save a new study, the system generates an experiment environment that is ready to pilot, edit, finalize, and deploy to collect data. Each study detail page includes an Overview, Modules, Module Data Elements, and Collected Data sections. The sections’ information is entered by TRS researchers when they create/edit a study or is generated when TRS researchers run configured modules (i.e., place a call) using the Participant Portal. MITRE software developers used a JSON structure to track study data elements due to its ability to save complex data as single values or arrays and its compatibility with a MongoDB database. The JSON structure enabled synchronize values with each step of the create/edit process and save study data elements to the MongoDB database. 

\begin{figure}[h]
\centering
\includegraphics[width=\textwidth]{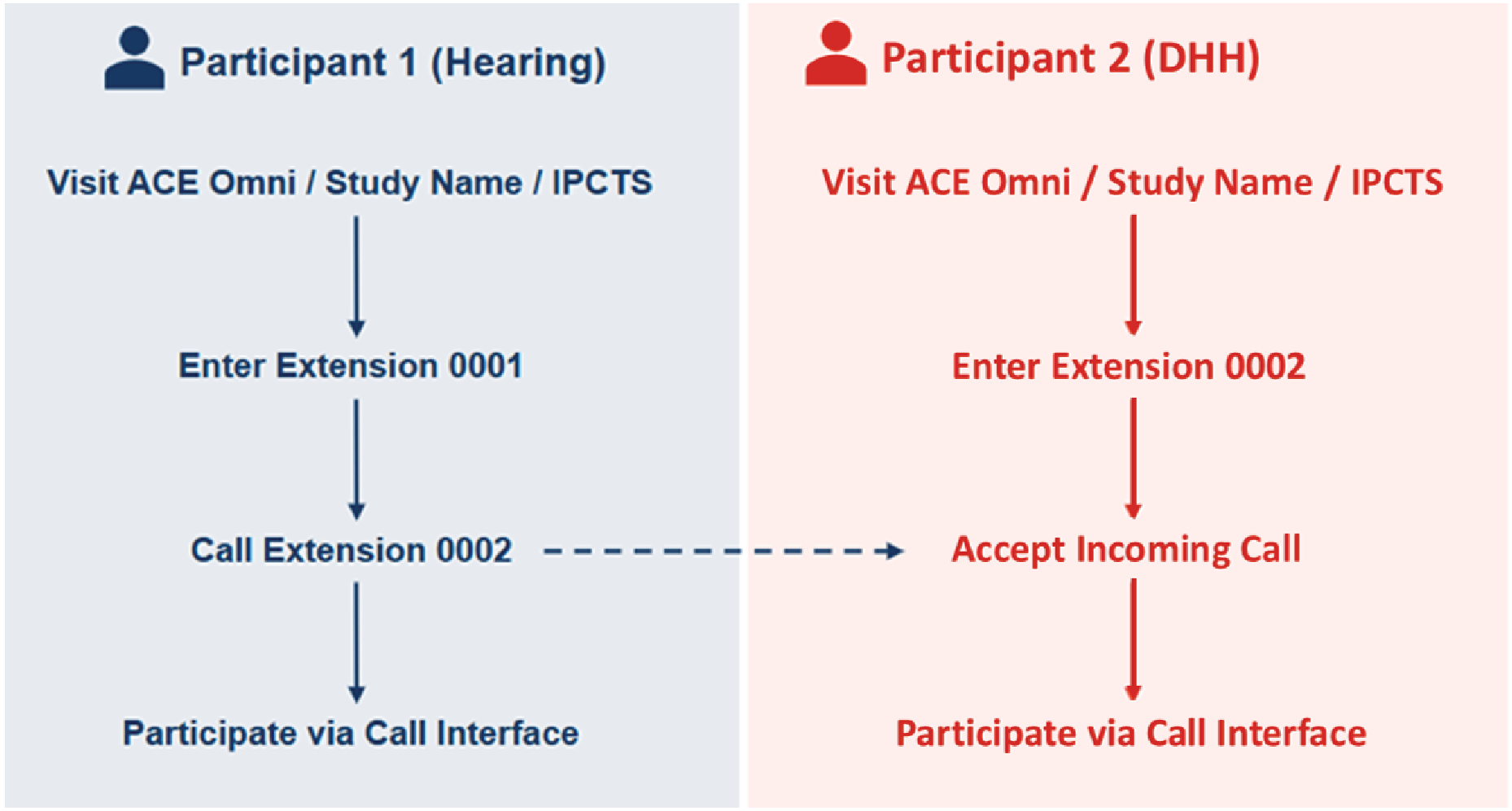}
\caption{Participant Portal workflows for a study using the IP CTS with ASR module.}\label{fig8}
\end{figure}

An Uniform Resource Locator (URL)/Quick Response code opens the Participant Portal displaying a unique experiment environment corresponding to a configured module within a study. In order for ACE Omni to direct participants to their respective call environments they must first enter their identifying extension in the Participant Portal, as shown in Fig.~\ref{fig8}. The participant with the lowest-numbered extension number (e.g., 0001) enters and calls the extension number of the participant designated to receive the call (e.g., 0002). When a call connects, ACE Omni automatically begins collecting data according to the selections made by TRS researchers in Step 5. TRS researchers can view a list of call metadata, including the date, start- and end-time, duration, and module configuration, and download call data in a CSV format in the Collected Data section of a study’s detail page.

\begin{figure}[h]
\centering
\includegraphics[width=\textwidth]{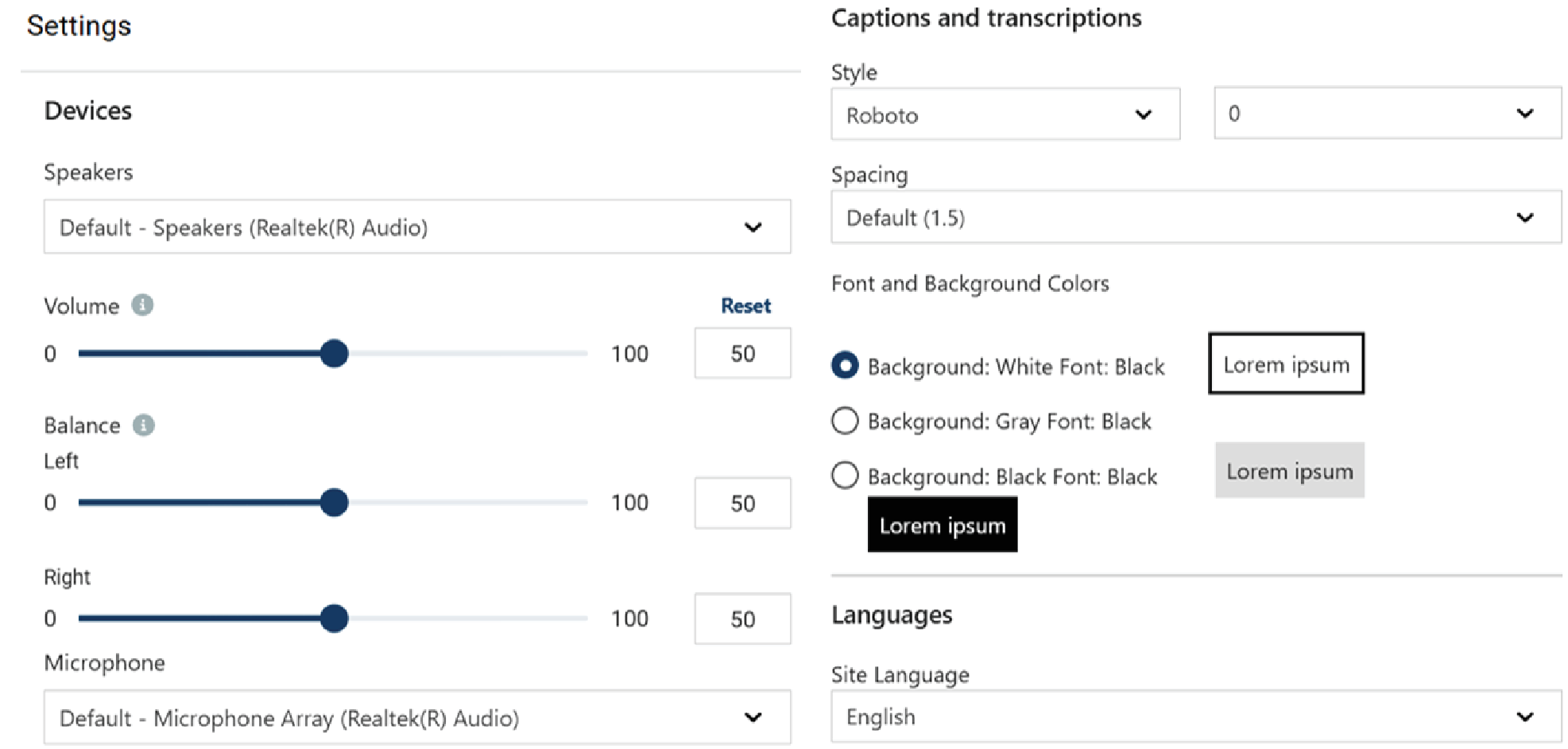}
\caption{Participant Portal accessibility features and other Settings for a study using the IP CTS with ASR module.}\label{fig9}
\end{figure}

When a call connects, TRS researchers and/or participants can manipulate and apply accessibility features and other settings via the Settings modal dialog on the participant interface. TRS researchers must consider how some options may conflict with variables TRS researchers may be studying. Accessibility features and other settings are specific to the module’s properties. For example, the ASR-based IP CTS module includes accessibility features for captions, which the VRS module does not include. An example of the Settings modal dialog using the ASR-based IP CTS module is shown in Fig.~\ref{fig9}. This Settings modal dialog includes three categories of accessibility features and settings: Devices, Captions and Transcriptions, and Languages. Devices enable TRS researchers and/or participants to select audio output and input devices, including speakers (e.g., internal device speakers versus external speakers), adjust system volume, adjust left/right balance of speakers, and select a microphone. The Captions and Transcriptions enable TRS researchers and/or participants to select the font style from a Google Fonts \cite{GoogleFonts} list, adjust font size, adjust font spacing, and select one of three font and background colors. The Languages option enables participants and/or TRS researchers to change the site language.

\section{Validation}\label{sec14}

MITRE conducted a study using ACE Omni, which was interested in investigating if connecting hearing devices (e.g., hearing aids or cochlear implants) directly to a call via Bluetooth improved audio perception (e.g., speech intelligibility) compared to listening to audio from an amplified phone speaker. The impacts of call interference (i.e., background noise) on audio perception were also explored. This validation assessed all ECF components. The following sections discuss the design of the study during pilot activities, during-study execution, and post-study execution and how lessons learned influenced changes to the study design or have implications for future work.

\subsection{Study Design}\label{sec15}

MITRE researchers collected data from individuals attending the 2023 Association of Late Deafened Adults conference (ALDAcon) for this study. MITRE has conducted numerous evaluations over the years and has identified venues, including conferences, that have individuals representative of the TRS user population. However, due to limited available time during conferences (i.e., attendees participated in various conference activities), it has become critical for MITRE researchers to employ strategies that guarantee successful data collection. To determine what personnel, strategies, and equipment are needed to ensure data collection, MITRE began by examining the task. 

\subsubsection{Task and Protocol}\label{sec16}

The study will begin with the arrival of the participant at Environment \#1 (described in Section \ref{sec17}). Researcher A will greet the participant, receive their consent form, and confirm their pre-evaluation survey completion, which they receive when they sign up to participate in the study. Paper surveys with large text are used to improve readability for the participants. Researcher A will then inform the participant that video and audio will be recorded during the study and receive their consent to proceed forward. The MITRE researchers and participant will introduce themselves to one another and Researcher A will provide the participant information about the purpose of the study, describe the tasks they will complete, and what equipment will be used. 

The two conditions tested in this study are hearing audio via computer speakers (intended to mimic audio from an amplified phone) and hearing audio via a Bluetooth connection of the participants’ hearing device(s) to their respective phones. For each condition, Researcher B will play the role of the other caller and play nine distinct pre-recorded Harvard sentences \cite{Rothauser1969} to the participant (not the same sentences across conditions). The pre-recorded sentences were altered using the Audacity software to overlay a babble noise (-5dB, 0dB, 7.5dB) on the sentence audio files to simulate background noise and vary difficulty in speech intelligibility. After each sentence is played the participant will be instructed to verbally repeat what they thought they heard and complete a post-sentence survey. The conditions and sentences will be randomized and counterbalanced across the participants. 

A practice call will be executed to confirm the participant understands the tasks they need to complete. Researcher A will always set up the participant call to ensure no issues arise with receiving the call from Researcher B. Before practicing an example pre-recorded sentence, Researcher A will confirm with the participant that they are able to hear Researcher B and the pre-recorded sentence. Once the practice call and practice post-sentence survey are complete the first condition will be executed. 

In the Bluetooth condition Researcher B places a call from their laptop to the participant receiving the call on their respective phone, while in the Amplified Phone condition the calls are placed and received on laptops. Researcher A will always state what condition is being conducted (e.g., Bluetooth Condition) followed by which pre-recorded sentence is being played (e.g., Sentence A). After the last post-sentence survey is completed by the participant, a post-condition survey is completed. Researcher A will terminate the call while the participant completes the post-condition survey and begin setting up the next call condition. The same first condition steps are executed for the second condition; however, upon completion of the post-condition survey a post-experiment survey follows. The evaluation is complete once the participant receives their payment. When the participant leaves Environment \#1, Researchers A and C will stop the recording devices, save the corresponding files, and set up for the next participant. At the end of each day the MITRE researchers will upload all recorded data to their corresponding folders, scan the surveys and upload them to their corresponding folders, and hardcode the survey responses into an aggregated Excel spreadsheet.

\subsubsection{Environments}\label{sec17}

Two MITRE researchers, one who led the evaluation (Researcher A) and another that took notes and provided aid to Researcher A (Researcher C), and the respective participant were in Environment \#1. Researcher B was in Environment \#2. The hardware and software utilized in Environment \#1 and \#2 consisted of the items listed in Table \ref{tab3}. All of the MITRE researchers followed scripts indicating who was responsible for what actions and provided instructions for the participants. 

\begin{table}[h]
\caption{Environment \#1 and \#2 hardware and software equipment.}\label{tab3}
\begin{tabular*}{\textwidth}{@{\extracolsep\fill}ll}
\toprule%
\textbf{Environment} & \textbf{Hardware and Software Equipment} \\
\midrule
\#1 & 1. A set of detachable speakers that connected to the laptop so that the \\
 & participant could adjust the audio level (important for the Amplified Phone \\
 & condition). \\
 & 2. A MITRE cellphone connected on a tripod that recorded the environment's \\
 & activities using the camera application. \\
 & 3. The participant's respective cellphone (used for the Bluetooth condition). \\
 & 4. Reading glasses in case participants forgot theirs and needed them to \\
 & proceed with the study. \\
 & 5. The participant's hearing device(s). \\
 & 6. Physical paper surveys. \\
 & 7. Pencils. \\
 & 8. Participant payment. \\
Both \#1 and \#2 & 9. A laptop used to receive and make ACE Omni calls, play the pre-recorded \\
 & sentences, and for recording the laptop's screen activities and audio in the \\
 & room (using Mac's built in screen recording tool). \\
 & 10. A MITRE cellphone that Researcher A and B used to communicate (the \\
 & MITRE researchers were on a separate call and used headphones to hear one \\
 & another in order to not interrupt the participant). \\
 & 11. Laptop and cellphone chargers. \\
 & 12. A hotspot to ensure stable internet connection. \\
 & 13. A spreadsheet that indicated the condition (Bluetooth or Amplified Phone) \\
 & and sentence order for each participant. \\
\botrule
\end{tabular*}
\end{table}

\subsection{Pre-Study Pilot Runs and Lessons Learned}\label{sec18}

\begin{figure}[h]
\centering
\includegraphics[width=\textwidth]{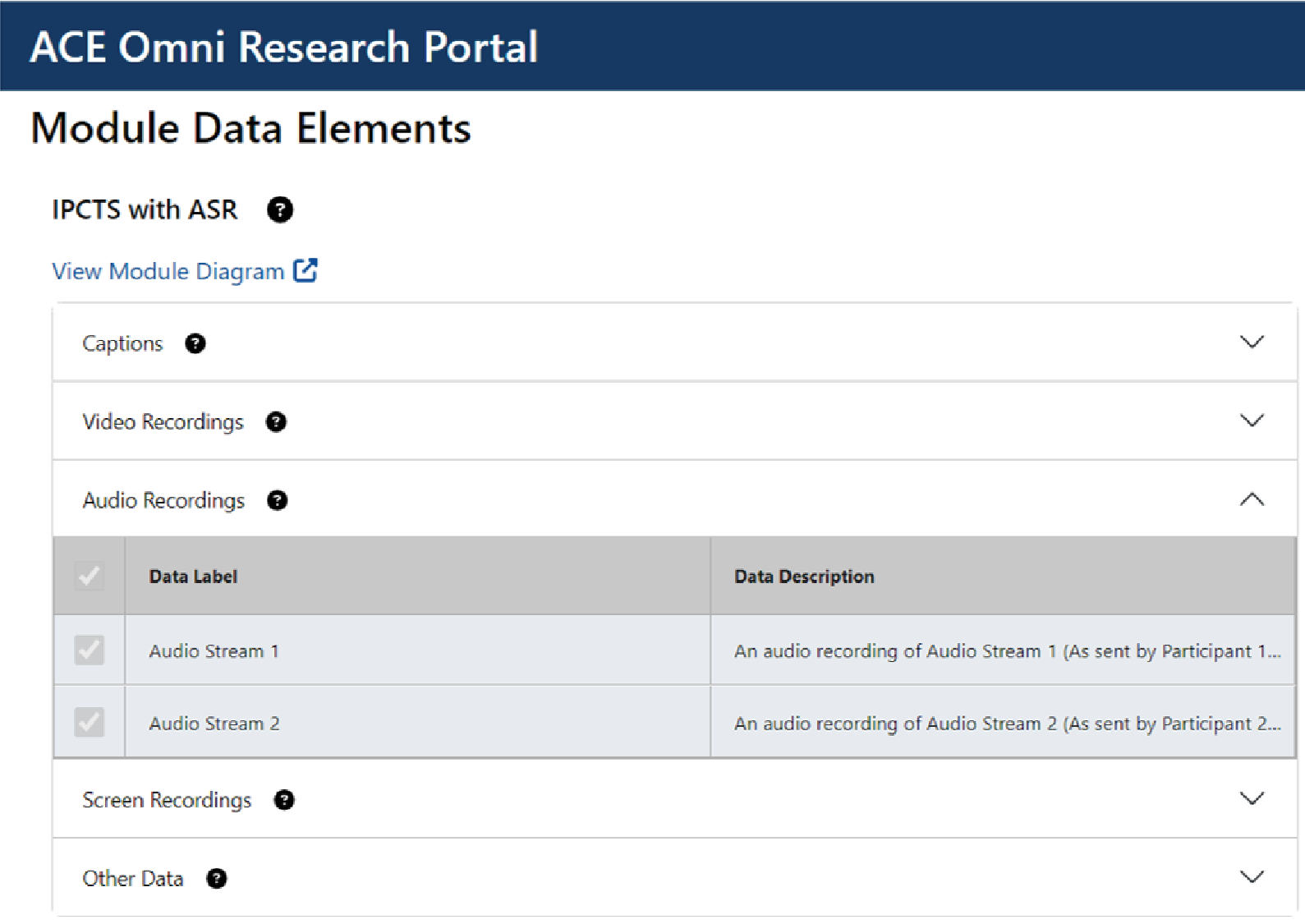}
\caption{ALDAcon 2023 study set up Step 5 selections.}\label{fig10}
\end{figure}

Prior to conducting the study various pilot tests were ran in MITRE labs with hearing MITRE researchers to validate the design of experiments. In this case, because there was no amplified phone module available in ACE Omni, the MITRE researchers used the IP CTS with ASR module and only selected the Audio Recordings option since audio data was the desired data type to collect (see Fig.~\ref{fig10}). 

\begin{figure}[h]
\centering
\includegraphics[width=\textwidth]{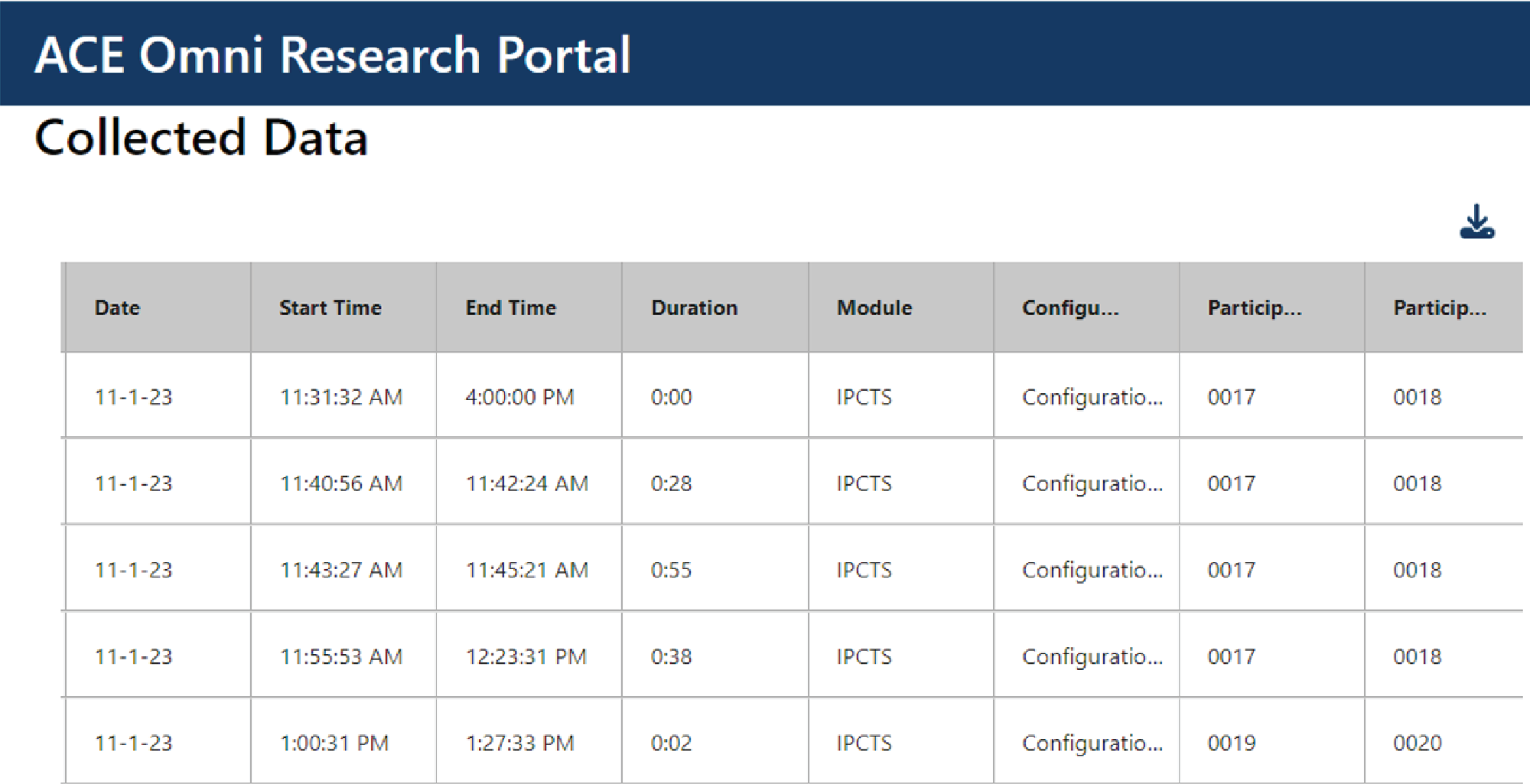}
\caption{ALDAcon 2023 example of data collected.}\label{fig11}
\end{figure}

Due to limited available time at ALDAcon the MITRE researchers only configured 18 calls anticipating a maximum of 18 participants in the study. Each configuration specified the callers (i.e., Researcher B and a participant) and their respective identification numbers (examples shown in Fig.~\ref{fig11}). 

During pilot testing three issues arose (see Table \ref{tab4}) that were not anticipated or accounted for during ACE Omni development or during the design of experiments. 

\begin{table}[h]
\caption{Issues experienced during piloting.}\label{tab4}
\begin{tabular*}{\textwidth}{@{\extracolsep\fill}ll}
\toprule%
\textbf{Issue Number} & \textbf{Description} \\
\midrule
\#1 & The inability for the MITRE researchers to place ACE Omni calls when using \\
 & some cellphone brands (e.g., iPhone versus Google). \\
\#2 & MITRE's secure firewall inhibited MITRE researchers to place ACE Omni \\
 & calls in locations other than the office (e.g., at their respective homes). \\
\#3 & ACE Omni calls were disrupted due to cellphone settings: if the screen time \\
 & out feature was triggered, when the screen locked, and if another call was \\
 & trying to patch through the cellphone's call application. \\
\botrule
\end{tabular*}
\end{table}

After further inspection of the failure reports during piloting, it was discovered that a cellphone’s operating system’s predefined applications, such as Safari, will always take precedence over other downloaded applications, such as Chrome which was used during ACE Omni development. Therefore, the back-end code was updated to consider various browsers to mitigate issue \#1. The MITRE researchers also implemented alternative calling methods (e.g., using the participant’s cellphone’s call application or using Microsoft Teams) into the experiment protocol to ensure the evaluation could continue, instead of downloading applications on participants’ cellphones. Since numerous recording devices were used the alternative calling methods data would still be collected. 
Initially the only IP addresses that had access to ACE Omni were those within MITRE locations. However, as pilot testing expanded beyond MITRE locations the developers needed to change the access to open for all. Usernames and passwords were implemented in order to mitigate the possibility of unauthorized people accessing ACE Omni. Additionally, the ACE Omni URL was not publicly advertised. This supported the MITRE researchers’ decision to handle call set up on behalf of the participants. These strategies addressed the challenges with issue \#2.

Two modifications were made to the protocol to mitigate issue \#3. First, additional information was provided during recruitment to participants that the MITRE researchers will need to temporarily alter the screen time out, lock screen, and do not disturb cellphone features on participant cellphones in order to execute ACE Omni calls effectively. This required participants to consider if and when they could participate, especially if they were expecting to receive calls. Secondly, instructions were added to the scripts instructing MITRE researchers to annotate what the original settings of the participant’s cellphone were, prior to changing settings, and to show the participants what the changes were on their cellphones. The changes to participants’ cellphones were only made prior to the practice call in the Bluetooth condition and were immediately changed back after the condition was completed. 

\subsection{Study Execution and Lessons Learned}\label{sec19}

Sixteen participants took part in the study. All of the participants were over 50 years of age, and 88\% of them reported having a severe hearing handicap. The participants used a wide range of hearing aids and cochlear implants, including: Starkey Evolv AI, Phonak Naida, Oticon OPN, GN Resound, Signia Pure 312X, Advanced Bionics Marvel CI, and Nucleus 7/8. While executing the study additional issues arose (see Table \ref{tab5}), distinct from those experienced during development and pilot testing. 

\begin{table}[h]
\caption{Issues experienced during study execution.}\label{tab5}
\begin{tabular*}{\textwidth}{@{\extracolsep\fill}ll}
\toprule%
\textbf{Issue Number} & \textbf{Description} \\
\midrule
\#4 & The incorrect assumption that participants understood and knew how to use \\
 & their hearing devices and applications to properly connect them via Bluetooth. \\
\#5 & Participants' cellphone applications had prioritization over ACE Omni. \\
\#6 & ACE Omni crashed because too many calls were placed during a given \\
 & timeframe. \\
\#7 & ACE Omni experienced audio challenges, such as call audio distortion (i.e., \\
 & anonymous censored voice) or no sound was heard. \\
\#8 & ACE Omni call connection challenges in Bluetooth condition. \\
\botrule
\end{tabular*}
\end{table}

There were instances, associated with issue \#4, when participants did not realize that their hearing devices were not connected via Bluetooth to their cellphones, which automatically terminated the Bluetooth condition to mitigate connectivity issues. There were also cases when participants did not know how to connect their hearing devices to their cellphones. Many participants mentioned that their audiologist initially set up the Bluetooth connection between their hearing devices and cellphone applications and rarely taught the participant how to connect their devices themselves. Some participants, who used hearing aids and cochlear implants, were unable to connect both devices via Bluetooth. This was likely due to the lack of bimodal wireless pairing, which has just recently become available. When hearing devices did not connect to cellphone applications, participants tended to restart their Bluetooth devices, which usually helped remediate connectivity issues. 

Some participants did not receive ACE Omni calls because of their cellphone’s applications (issue \#5), such as Innocaption or Thrive Hearing Control, which were prioritized over ACE Omni. The ACE Omni audio was routed through those applications and participants could not hear the MITRE researcher. Calls were therefore placed through the participant’s call application (an alternative calling method) and the audio recording methods were used to collect data. 

Upon investigation of issue \#6 it was determined that the MITRE researchers needed to modify their protocol by closing the browser to reset the environment, which ensured previous call connections were cleared before starting a new call. The MITRE researchers also cleared the cache data from the previous day each morning to mitigate this issue. Numerous long duration calls placed during the study execution caused the issue to arise. Development and pilot testing missed these situations because a limited number of call extensions were used, and evaluation runs were shorter. 

Investigations into issue \#7 identified various factors and interactions that could attribute to the challenges; however, due to time constraints, short-term mitigation strategies were implemented to ensure study execution. The MITRE researchers began by reconnecting to the hotspot Wi-Fi in case of network instability issues, followed by restarting the ACE Omni call. If both strategies did not yield positive results, the MITRE researchers executed an alternative calling method to continue with the study. Post-study, it was determined that audio recorded with an iPhone was recorded with a 48000 or 44100 Hz sample rate, but the ACE Omni file data was 16000 Hz. The mismatch in sample rate caused the audio to playback with a distorted effect. During data analysis, distorted ACE Omni files were found even if distortion was not present during study execution calls. Therefore, the MITRE researchers relied on the alternative calling method files to analyze data. 

Investigations into issue \#8 yielded similar conclusions to issue \#4; therefore, similar mitigation strategies were leveraged to continue study execution.  

\subsection{Discussion and Future Work}\label{sec20}

ACE Omni was created to help reduce challenges TRS researchers experience when setting up TRS experiment environments, which includes setting up TRS solutions and the data collection methods. The results of the piloting and study validation activities support that ACE Omni was successful at helping MITRE researchers conduct their experiment, despite experiencing some issues. The majority of issues were not rooted in ACE Omni, but were challenges associated with connectivity (ECF Channel), the participant’s cellphone (ECF Channel), and the participants (TRS Caller). Issues experienced during piloting were not experienced during study execution. MITRE is currently proposing future work to further develop ACE Omni to address issues that were not fixed and to integrate the recommendations from Section \ref{sec22}.

\subsubsection{Issues Experienced During Pilot Runs and Study Execution}\label{sec21}

The connectivity issues that arose were related to secure firewalls and cache build up, which inhibited ACE Omni calls to be placed. Changing ACE Omni to include usernames and passwords and changing research protocols, such as closing browsers after every ACE Omni call and clearing the cache at the end of each research day, remediated these issues. The challenges experienced with participants’ cellphones were related to cellphone settings and applications. Screen time out, screen lock, and receiving phone calls via the cellphone’s call application were settings that disrupted ACE Omni calls, while particular applications, such as captioning or hearing device applications, restricted ACE Omni calls to be placed altogether. Recruitment and research protocol modifications were made to inform participants that MITRE researchers needed to change some settings on their cellphones during the study to ensure study execution was uninterrupted. However, more complex settings specific to captioning and hearing device applications were not changed by the MITRE researchers. Manipulating the participants’ hearing devices’ software and hardware were avoided to mitigate the risk of potentially negatively impacting the participants’ ability to hear if changes were made. Therefore, alternative calling methods were implemented into the research protocol to ensure data was collected from the participants, although these alternatives did not capture the participants’ real life call experience with their preferred devices. 

The final challenge, not associated with ACE Omni, was MITRE researchers assuming that the participants understood how to use their hearing devices and their respective applications to connect them via Bluetooth. Since the MITRE researchers wanted to avoid manipulating participants hearing devices’ software and hardware, the participants could only complete the Amplified Phone condition. This decision resulted in only half the data being collected for those particular participants. Future studies could consider alternative strategies to mitigate this risk, such as employing audiologists to participate in the study evaluation. However, it would be imperative for TRS researchers to collect as much information from the participants during recruitment to determine if the audiologist was familiar with the identified hearing devices’ software and hardware in case issues arose during study execution. It is unlikely that any single audiologist would be familiar with every type of available hearing device software and hardware; therefore, employing multiple audiologists or excluding participants with unfamiliar hearing device software and hardware may need to be considered.  

Two ECF Channel issues occurred and it was unclear if the issues were related to ACE Omni or a combination of ACE Omni and the participants’ cellphones. The first challenge were audio issues during and post-study execution. There were instances when the audio was distorted or no sound was heard during execution, which required employing a variety of methods to continue with the study. There were also instances when no audio issues were experienced during study execution, but the recorded data was distorted. MITRE software developers investigated these issues; however, no single factor was identified as the cause of the issues. The second issue was related to ACE Omni call connections. On occasion the MITRE researcher attempted to place an ACE Omni call in the Bluetooth condition, but the call was never patched through. Similar investigations were conducted by MITRE software developers, but no conclusions were made. Further investigations are needed to remediate these two issues.

The one issue that was experienced during piloting that was directly related to ACE Omni (ECF Channel) was the inability for MITRE researchers to place ACE Omni calls when using non-Android cellphones. During development, only Android cellphones were used to test ACE Omni. When the MITRE software developers investigated what may be causing the issue they identified back-end code that needed to be modified to consider a variety of cellphone and browser types to mitigate the issue. Once the code was modified the issue did not occur again. This issue exemplifies the necessity for TRS researchers to communicate with developers during the development stage what future use cases need to be considered to ensure that a variety of software and hardware devices can leverage ACE Omni.

\subsubsection{Recommendations to Improve ACE Omni}\label{sec22}

Overall ACE Omni worked well, considering only one of eight issues was explicitly associated with the ACE Omni system. However, there are recommendations (see Table \ref{tab6}) to improve ACE Omni in future work if a similar type of study is conducted. Some of these recommendations (\#1-3) emerged from the outputs in Section \ref{sec7} but were not integrated into ACE Omni to ensure that core system functionality was operational for the validation study. Recommendation \#4 was derived from the lessons learned during the validation study and demonstrates how a TRS researcher need was discovered during data analysis. Each recommendation is described with a potential solution; however, these recommendations require validation from future evaluations. 

\begin{table}[h]
\caption{ACE Omni improvement recommendations.}\label{tab6}
\begin{tabular*}{\textwidth}{@{\extracolsep\fill}ll}
\toprule%
\textbf{Recommendation Number} & \textbf{Enable TRS Researchers To:} \\
\midrule
\#1 & Introduce distortion or corruption to audio streams in ACE Omni. \\
\#2 & Build and embed surveys into ACE Omni. \\
\#3 & Type or append notes during or post-ACE Omni calls to ensure \\
 & accurate note taking. \\
\#4 & Collect transcription data of ACE Omni call audio. \\
\botrule
\end{tabular*}
\end{table}

The MITRE researchers in the validation study relied on Windows Media Player to play the sentence audio files, which required splitting their laptop screen between the browser and the participant folders. Providing the ability to distort/corrupt audio streams in ACE Omni via module configuration (recommendation \#1) can help reduce the need for TRS researchers to navigate various applications during calls. Since ACE Omni treats different data streams of modules (e.g., audio streams of the IP CTS with ASR module) as discrete items with sets of properties that can be configured in Step 4: Configure Components, additional properties could be added to audio streams to enable TRS researchers to distort or corrupt them. One property that could be added to distort audio streams is background noise. This property can be manipulated by enabling TRS researchers to upload an audio file that plays over the audio stream and specify how long it should play for and if it should or should not loop. TRS researchers can control when an audio file starts/pauses/resumes playing via an onscreen affordance on each participant’s call screen (e.g., a ‘Start/Pause Audio File’ button). To corrupt audio streams, a property could be added that enables TRS researchers to simulate packet drops (i.e., when units of data being transmitted over a network and its metadata fails to reach their intended destination). Packet drops tend to make the audio sound choppy since chunks of audio are being ‘dropped’. This property can be manipulated by enabling TRS researchers to select the duration of the packet drop, and if the packet drop should be at regular or random intervals. Like audio files, TRS researchers can start/pause/resume packet drops via onscreen affordances. TRS researchers can use background noise and simulated packet drops individually or simultaneously, depending on the research conditions. 

The MITRE researchers used paper surveys to collect data, which resulted in a high number of materials to execute the study and a large duration of time to manually input paper survey data into Excel spreadsheets. Embedding in-call surveys into ACE Omni (recommendation \#2) would help reduce materials, time, and complexity of tracking data. One way that surveys could be embedded into ACE Omni calls is by implementing a separate survey building tool, where TRS researchers can create surveys and then specify how/where to present them to the participants. The survey building tool could be implemented within the Create New Study process and enable TRS researchers to add surveys to individual module configurations. Similar affordances as those mentioned for recommendation \#1 can be provided to TRS researchers to display the survey, such as a ‘Start Survey’ button. 

Displaying surveys to participants in the same medium during calls, however, presents unique challenges that must be considered. If a survey was to be presented to a participant, how should it be displayed? Where should it be displayed? The same considerations that many videoconferencing platform designers need to contemplate for their users are applicable in this situation. For example, participants that use captions must have access to those captions and they must be visible. However, displaying a survey (similar to that of sharing a document during a videoconference call), whether it’s on the same ACE Omni Participant Interface or in an additional pop-up window, will take up screen space potentially affecting captions. These considerations, and more, are critical to contemplate when developing a design solution that is usable by participants and meets researcher needs.

Research notes could be directly stored in ACE Omni rather than other note taking mechanisms (recommendation \#3), which would reduce the number of systems/applications TRS researchers need to use to collate notes. To accommodate this recommendation, the way ACE Omni stores data on the backend can be adjusted. Instead of simply storing all call data in one folder, call data can be stored in a folder that contains participant folders. This would enable any notes added during or post-ACE Omni calls for specific participants to be organized by call participant. Additionally, note-taking affordances would need to be added to the front end, including ways for TRS researchers to add notes during calls using their own devices as well as in the Collected Data section of a study’s detail page. One concept to enable TRS researchers to take notes during calls is a moderation panel, which enables TRS researchers to manage study sessions in real time. TRS researchers can open a study’s moderation panel on that study’s detail page and the panel can link to individual calls when the call connects. In the panel, TRS researchers can create a note and select whether the note is general (i.e., applies generally to the call) or relevant to a specific participant. When TRS researchers are done using the moderation panel, they can close the panel. For each row of call data in the Collected Data table, affordances (e.g., ‘view notes’ buttons) can be added to a ‘Notes’ column that enables TRS researchers to view notes for that call. To add notes post-ACE Omni call, an ‘Add Notes’ button can be added to that same ‘Notes’ column that, when pressed, opens a modal dialog with options for appending notes to that call.

Transcripts document verbal communication from a call and are stored typically as a text file, while captions show written versions of verbal communication to participants live during a conversation. Both transcripts and captions can output the same data; however, they can deviate depending on what ASRs or CAs are being leveraged. To accommodate recommendation \#4, an additional option in Step 5 can be added that enables TRS researchers to turn on/off transcript collection. The option would be toggled on by default, similar to other data collection options, but TRS researchers could turn it off depending on their data collection needs. The option should be paired with a description to help differentiate it from captions. If transcript collection is turned on, transcripts will be collected for all calls for that study and can be incorporated into the call file that researchers download in the Collected Data section.

\subsection{Conclusion}\label{sec23}

Mechanisms to improve the process of conducting TRS research are needed to promote more TRS research. The results of TRS research are used to inform how the industry can close functional equivalence gaps and develop TRS rules and regulations; therefore, it is pertinent to reduce the barriers-of-entry to conduct TRS research. To help mitigate TRS research challenges, MITRE developed the ACE Omni platform using a number of design methodologies, information gathering strategies, and work products. ACE Omni underwent validation activities including in-lab pilot sessions and an in-field study evaluation. The majority of challenges that arose during the validation activities were associated with connectivity issues, challenges with the participant’s cellphone, and a lack of participants’ knowledge of their hearing devices’ hardware and software. Only one issue, that occurred during the in-lab piloting activities, was directly related to ACE Omni; the inability for MITRE researchers to place ACE Omni calls when using non-Android cellphones. All of the issues and challenges experienced were addressed rapidly after their identification and through changes to ACE Omni and/or the study protocol. ACE Omni has proven to be a valuable research tool and was successfully leveraged in a TRS study. Future ACE Omni development can include adding module properties, enabling TRS researchers to manage studies in real time on their own devices, and incorporating additional data collection options. To MITRE’s knowledge, ACE Omni is the only research platform that supports TRS research.

\backmatter

\section*{Declarations}

\subsection{Funding}

This work was supported by the Federal Communications Commission under Federal Contract \#75FCMC18D0047, Task Order \#273FCC19F0114 and the MITRE Corporation's Research Enablement and Augmentation Program. ©2025 THE MITRE Corporation. All Rights Reserved. Approved for Public Release; distribution unlimited; case \#25-1488.

This ACE Omni software was produced for the U. S. Government under Contract Number 75FCMC18D0047/75FCMC23D0004, and is subject to Federal Acquisition Regulation Clause 52.227-14, Rights in Data-General.  

No other use other than that granted to the U. S. Government, or to those acting on behalf of the U. S. Government under that Clause is authorized without the express written permission of The MITRE Corporation. 

For further information, please contact The MITRE Corporation, Contracts Management Office, 7515 Colshire Drive, McLean, VA  22102-7539, (703) 983-6000.  
© 2025 The MITRE Corporation.

\subsection{Author Contribution}

The authors contributed equally to this work. Harrison Bourikas, Karina Roundtree, and Vincent Ybarra supervised the work.




\bibliography{sn-bibliography}

\end{document}